\documentclass[preprint,prc,showpacs,preprintnumbers,amsmath,amssymb]{revtex4-1}
\usepackage{graphicx,latexsym,amssymb,amsmath,color,multirow,mathrsfs,booktabs, 
ifpdf}
\usepackage{dcolumn}% Align table columns on decimal point
\usepackage{bm}% bold math
\usepackage{longtable}

\def\pA{proton-nucleus\ }
\def\heA{$^3$He-nucleus\ }

\def\cx{charge-exchange\ }
\def\phe6{$^6$He+$p$\ }
\def\h6pn{$p(^6$He,$^6$Li$_{\rm IAS})n$\ }
\def\pn{$(p,n)$\ }
\def\he3t{$(^3$He,$t)$\ }

\def\xpb{$^{208}$Pb($^3$He,$t)^{208}$Bi$_{\rm IAS}$\ }

\def\pphe6{$p(^6$He,$^6$He)$p$\ }
\def\p2he6{$p(^6$He,$^6$He$^*)p$'\ }
\def\nli6{$^6$Li$^*+n$\ }

\begin{document}
\title{Charge-exchange scattering to the isobaric analog state at medium energies
as a probe of the neutron skin\footnote{\bf Dedicated to the memory of Ray Satchler}}
\author{Bui Minh Loc$^{1,2}$}
\author{Dao T. Khoa$^1$}
\author{R.G.T. Zegers$^3$}
\affiliation{$^1$Institute for Nuclear Science and Technology, VINATOM \\
179 Hoang Quoc Viet Rd., Hanoi, Vietnam.\\
$^2$ University of Pedagogy, Ho Chi Minh City, Vietnam. \\
$^3$ National Superconducting Cyclotron Laboratory and \\
Department of Physics and Astronomy, \\
Michigan State University, East Lansing, MI 48824-1321. }
\begin{abstract}{The \cx \he3t scattering to the isobaric analog 
state (IAS) of the target can be considered as ``elastic" scattering 
of $^3$He by the isovector term of the optical potential (OP) 
that flips the projectile isospin. Therefore, the accurately measured 
\cx scattering cross-section for the IAS can be a good probe of the isospin
dependence of the OP, which is determined exclusively within the folding
model by the difference between the neutron and proton densities and isospin 
dependence of the nucleon-nucleon interaction. Given the neutron skin 
of the target is related directly to the same density difference, it can be 
well probed in the analysis of the \cx \he3t reactions at medium energies 
when the two-step processes can be neglected and the $t-$matrix interaction 
can be used in the folding calculation. For this purpose, the data of the 
\he3t scattering to the IAS of $^{90}$Zr and $^{208}$Pb targets at 
$E_{\rm lab}=420$ MeV have been analyzed in the distorted wave Born approximation 
using the double-folded \cx form factor. The neutron skin deduced for these 
two nuclei turned out to be in a good agreement with the existing database.}
\end{abstract}
\date{\today}
%\pacs{25.55.Ci; 21.10.-k; 24.10.Ht; 24.10.Eq}
\maketitle
	
The neutron skin thickness determined as the difference between the neutron
and proton (root mean square) radii,
\begin{equation}
 \Delta R_{np}=\langle r^2_n\rangle^{1/2}-\langle r^2_p\rangle^{1/2},
\label{e1}
\end{equation}
was found by numerous structure studies to be strongly correlated with 
the slope of the symmetry energy of nuclear matter, i.e., the
density dependence of the symmetry energy \cite{Hor14,Tsang12}, which in
turn is a key quantity for the determination of the equation of state  
of the neutron-rich nuclear matter. As a result, an accurate determination 
of the neutron skin has become an important research object of different
nuclear reaction and structure studies (see a more detailed overview in 
Ref.~\cite{Tsang12}). 

Although it is straightforward that we need to choose a well-defined and 
closely related to the neutron skin quantity that can be measured with high 
precision, the sensitivity of such experimental data to the neutron skin
is often indirect and model dependent. Usually, a correlation between
the neutron skin and such an experimental quantity is carefully investigated 
in some structure model using a realistic choice of the effective 
nucleon-nucleon ($NN$) interaction, and a conclusion on the neutron skin 
of the considered nucleus is then drawn. A recent example is the study of the 
electric dipole polarizability $\alpha_D$ of $^{208}$Pb \cite{Roca13} based on 
both the microscopic random phase approximation approach and macroscopic 
droplet model, which gives $\Delta R_{np}\approx 0.165$ fm (with a total 
uncertainty around 25\%) for this nucleus.      
Another famous attempt is the Lead Radius Experiment (PREX) at the Thomas 
Jefferson laboratory \cite{prex}, where one has measured the parity-violating 
electron scattering on the $^{208}$Pb target and deduced in a rather 
model-independent way the neutron radius based on a larger weak charge 
of the neutron compared to that of the proton. As a result, the PREX data  
suggested a neutron skin $\Delta R_{np}\approx 0.33^{+0.16}_{-0.18}$ fm for 
$^{208}$Pb. 

It is clear from Eq.(\ref{e1}) that the neutron skin is directly related
to the difference between the neutron and proton densities, $\rho_n-\rho_p$,
that is also known as the nuclear isovector (IV) density. The \cx reactions 
\pn or \he3t are the well-known probes of different IV excitations, 
like the isobaric analog state (IAS), Gamow-Teller states and spin-dipole 
resonances. The IAS of the $(Z+1,N-1)$-nucleus has the same structure as 
the ground state (g.s.) of the $(Z,N)$-target except for the replacement 
of a neutron by a proton, and its excitation energy is, therefore, almost
equal to the Coulomb energy of the added proton. The two IAS's are members 
of an isospin multiplet which have similar structures and differ only in 
the orientation of the isospin $\bm{T}$. Therefore, the \pn or \he3t reaction 
to the IAS can be approximately considered as an ``elastic" scattering 
process, with the isospin of the incident proton or $^3$He being flipped 
\cite{Dri62,Sat64,Sat69}. In such a picture, the \cx isospin-flip 
scattering to the IAS is naturally caused by the IV part of the optical 
potential (OP), expressed in the following Lane form \cite{La62} 
\begin{equation}
 U(R)=U_0(R)+4U_1(R)\frac{{\bm t}{\bm T}}{aA}. \label{e2}
\end{equation}
Here $t=1/2$ is the isospin of the projectile and $T$ is that of the target 
with mass number $A$, $R$ is the radial separation between the projectile 
and target, $a=1$ and 3 for the incident proton and $^3$He, respectively. 
The second term of Eq.~(\ref{e2}) is the \emph{symmetry term} 
of the OP, with $U_1$ known as the Lane potential that contributes to both 
the elastic scattering and \cx scattering to the IAS \cite{Sat69}. The 
empirical IV term of the \pA or \heA OP in the Woods-Saxon form has been 
first used by Satchler {\it et al.} \cite{Dri62,Sat64} as the \cx form factor 
(FF) to describe the \pn or \he3t scattering to the IAS within the 
distorted wave Born approximation (DWBA). 

In the standard isospin representation \cite{Sat83}, the target nucleus $A$ 
and \emph{isobaric analog nucleus} $\tilde{A}$ can be referred to as the 
isospin states with $T_z=(N-Z)/2$ and $\tilde{T_z}=T_z-1$, respectively.  
If we denote the state formed by adding proton or $^3$He to $A$ as 
$|aA\rangle$ and that formed by adding a neutron or triton to $\tilde{A}$ as 
$|\tilde{a}\tilde{A}\rangle$, then the DWBA \cx FF for the \pn or \he3t scattering 
to the IAS is readily obtained \cite{Sat83} from the transition matrix element 
of the OP (\ref{e2}) as
\begin{equation}
 F_{\rm cx}(R)=\langle\tilde{a}\tilde{A}|4U_1(R)\frac{{\bm t}{\bm T}}{aA}|
 aA\rangle=\frac{2}{aA}\sqrt{2T_z}U_1(R).
 \label{e3}
\end{equation}

The nucleon OP has been studied over the years, with several global sets 
of the OP parameters established from the extensive optical model (OM) 
analyses of the elastic nucleon scattering. Because the high-precision \pn 
data are not available for a wide range of target masses and proton energies, 
the IV term of the nucleon OP has been deduced \cite{BG69,Va91,Kon03} mainly 
from the OM studies of the elastic proton and neutron scattering from the same 
target and energy, where the IV term of the OP (\ref{e2}) has the same strength, 
but opposite signs for proton and neutron. Only in few cases has the Lane 
potential $U_1$ been deduced from the DWBA studies of \pn scattering to the 
IAS \cite{Car75,Jon00}. With the Coulomb correction properly taken into account 
\cite{devito}, the phenomenological Lane potential has been shown to account 
quite well for the \pn scattering to the IAS. However, a direct connection 
of the OP to the nuclear density can be revealed only when the OP is obtained 
microscopically from the folding model calculation \cite{Kho02,Kho07,Kho96,Kho14}. 

The isospin dependence of the \heA OP has been less investigated. 
Moreover, the recent \emph{global} OP for $^3$He and triton \cite{Pang09} 
accounts fairly well for the elastic scattering data using a purely 
\emph{isoscalar} real OP (with a slight dependence of the imaginary OP 
on the neutron-proton asymmetry). Therefore, one can learn more about the 
IV part of the \heA OP only in the study of the \cx \he3t reactions. 
In fact, the \he3t scattering to the IAS has been studied in the 
DWBA with the FF obtained from a single-folding calculation, using the 
effective (isospin-dependent) $^3$He-nucleon interaction and microscopic 
nuclear transition density for the IAS excitation \cite{Werf1,Werf2}. 
The present work is our attempt to study the \he3t scattering to the IAS 
based on the Satchler's prescription (\ref{e3}). Thus, the FF of the 
\he3t scattering to the IAS can be obtained from the double-folding model 
(DFM) \cite{Kho96,Kho14} in the following compact form     
\begin{eqnarray}
 F_{\rm cx}(R)=\sqrt{\frac{2}{T_z}}\int\int[\Delta\rho_1(\bm{r}_1)
 \Delta\rho_2(\bm{r}_2)v^{\rm D}_{01}(E,s)+
\Delta\rho_1(\bm{r}_1,\bm{r}_1+\bm{s}) \nonumber\\
\times\Delta\rho_2(\bm{r}_2,\bm{r}_2-\bm{s})
v^{\rm EX}_{01}(E,s)j_0(k(E,R)s/M)]d^3r_1d^3r_2,\label{e4} 
\end{eqnarray}
where $v^{\rm D}_{01}$ and $v^{\rm EX}_{01}$ are the direct and exchange
parts of the isospin-dependent term of the central $NN$ interaction;  
$\Delta\rho_i(\bm{r},\bm{r}')=\rho^{(i)}_n(\bm{r},\bm{r}')-
\rho^{(i)}_p(\bm{r},\bm{r}')$ is the IV density matrix of the 
$i$-th nucleus, which gives the local IV density when $\bm{r}=\bm{r}'$;
$\bm{s}=\bm{r}_2-\bm{r}_1+\bm{R}$, and $M=aA/(a+A)$.
The relative-motion momentum $k(E,R)$ is given self-consistently
by the real OP at the distance $R$ (see details in Ref.~\cite{Kho96,Kho14}).
In the limit $a\to 1$ and $\Delta\rho_1\to 1$, the integration over $\bm r_1$
disappears and (\ref{e4}) is reduced to a single-folded expression for
the FF of the \pn scattering to the IAS \cite{Kho14}.  
Because the energies of the analog states are separated approximately by the 
Coulomb displacement energy, the \cx scattering to the IAS has a nonzero 
$Q$ value. To account for this effect, the double-folded FF (\ref{e4}) 
is evaluated at the energy of $E=E_{\rm lab}-Q/2$, midway between the 
energies of the incident $^3$He and emergent triton, as suggested
by Satchler {\sl et al.} \cite{Sat64}. 

At the incident energies of $100\sim 200$ MeV/nucleon the impulse 
approximation is reasonable, and an appropriate $t$-matrix parametrization 
of the free $NN$ interaction can be used in Eq.~(\ref{e4}). 
Following the DWBA analysis of the \he3t reaction at the same energy 
to study the Gamow-Teller excitations \cite{Zer07,Per11}, we have used the in the 
present work the nonrelativistic $t$-matrix interaction suggested by 
Franey and Love \cite{Lo81,Fr85} based on the experimental $NN$ phase shifts.
Thus, the isospin-dependent direct and exchange parts of the central $NN$ 
interaction are determined from the singlet- and triplet even (SE,TE) and 
odd (SO,TO) components of the local $t$-matrix interaction \cite{Fr85} as
\begin {equation}
 v^{\rm D(EX)}_{01}(s)=\frac{k_ak_A}{16}[-3t_{\rm TE}(s)+t_{\rm SE}(s)
 \pm 3t_{\rm TO}(s)\mp t_{\rm SO}(s)].
 \label{e5}
\end {equation}
Here $k_a$ and $k_A$ are the energy-dependent kinematic modification factors 
of the $t$-matrix transformation from the $NN$ frame to the $Na$ and $NA$
frames, respectively, which are given by Eq.~(19) of Ref.~\cite{Lo81}. 
The explicit (complex) strength of the \emph{finite-range} $t$-matrix 
interaction (\ref{e5}) is given in terms of four Yukawa functions \cite{Fr85}. 
We note also that at medium energies, the two-step processes like 
($^3$He,$\alpha)(\alpha,t)$ or ($^3$He,$d)(d,t)$ are negligible and the 
direct \cx process is dominant, which allows one to deduce accurately the 
strength of the Fermi or Gamow-Teller transitions \cite{Zer07,Per11}. 

Another important nuclear structure input to the folding integral (\ref{e4})  
are the neutron and proton g.s. densities of the $^3$He projectile
and target nucleus. In the present work, we have used the neutron and proton 
densities of $^3$He given by the microscopic three-body calculation 
\cite{Nie01} using the Argonne $NN$ potential. For the $^{90}$Zr and 
$^{208}$Pb targets we have used the empirical neutron and proton densities 
deduced from the high-precision elastic proton scattering at 800 MeV by 
Ray {\it et al.} \cite{Ray78,Ray78c}. These densities are given in the 
analytic form so that one can slightly adjust the radial parameter of
the neutron density to the best DFM + DWBA description of the \cx \he3t 
scattering data under study, and determine the corresponding neutron skin.
For the comparison, we have also used the microscopic nuclear g.s. densities
given by the Hartree-Fock-Bogoliubov (HFB) calculation \cite{Gr01} using the
realistic Skyrme interaction and taking into account the continuum. These
HFB densities have been used earlier in the folding model analysis \cite{Kho04}
to study the total reaction cross sections measured at medium energies 
for the unstable nuclei.

The differential cross section of the \cx Fermi transition to the IAS, 
with $\Delta L=\Delta S=0$, and $\Delta T=1$, is known to peak at the
zero scattering angle. The absolute differential cross sections measured at the 
forward scattering angles ($\theta_{\rm lab}=0^{\circ}-2.5^{\circ}$) for the 
$^{90}$Zr,$^{208}$Pb($^{3}$He,$t$) reactions to the IAS were available 
from a previous study \cite{Zer07}, performed at the Grand Raiden Spectrometer 
at the Research Center for Nuclear Physics in Osaka. The $^{3}$He beam energy 
was 420 MeV. Further details of the experiment can be found in Ref.~\cite{Zer07}, and 
references therein. The uncertainty in the extracted absolute differential cross 
sections was about 10\% and predominantly related to the uncertainty in the current
integration of the unreacted beam in a Faraday cup. The $^{90}$Zr and $^{208}$Pb 
targets were isotopically pure ($>99\%$).

\begin{figure}[bht] \vspace*{-1cm}\hspace{-1cm}
\includegraphics[width=1.0\textwidth]{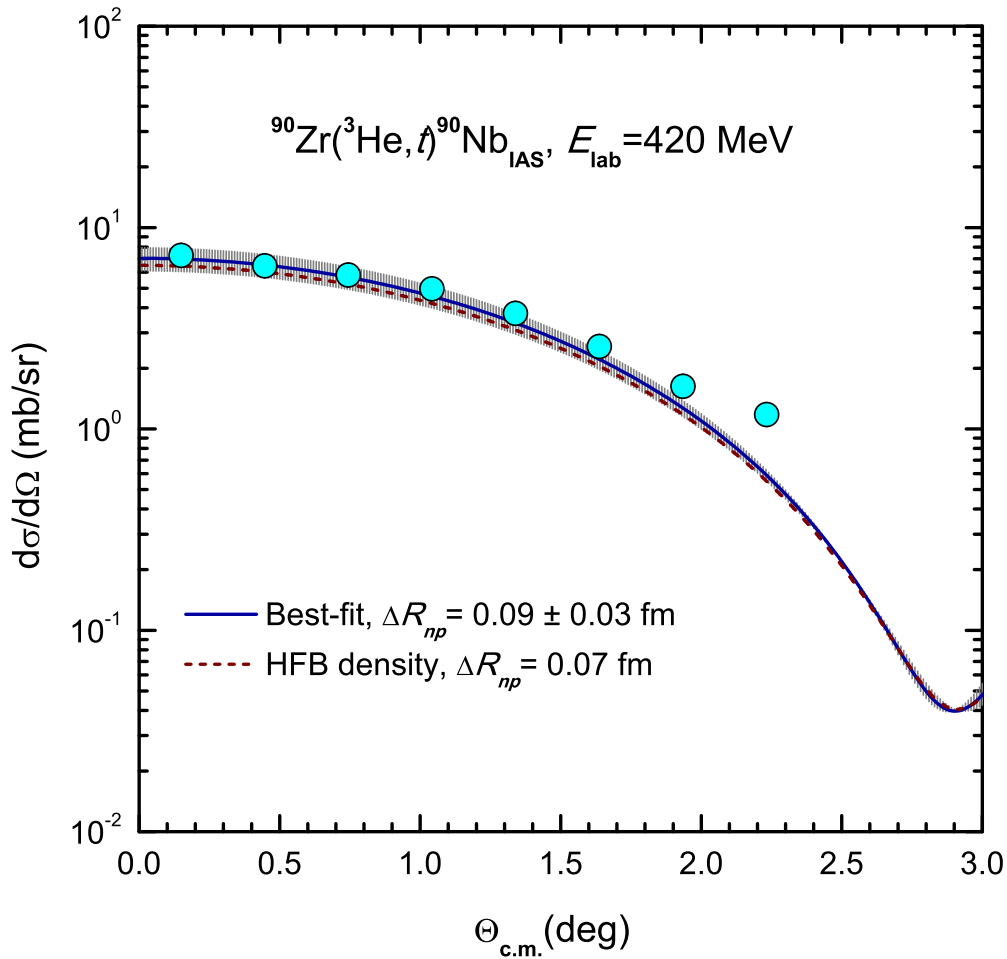}\vspace*{-1cm}
 \caption{DWBA description of the \he3t scattering to the IAS of the $^{90}$Zr
target given by the \cx FF (\ref{e4}) based on the empirical IV density by Ray 
{\it et al.} \cite{Ray78}, adjusted by the best DWBA fit to the data.
The error of the (best-fit) neutron skin was determined to account for the 
experimental uncertainty around 10\% of the absolute differential cross 
section measured at the most forward angles (the hatched area). The dash 
curve is the prediction given by the microscopic HFB density \cite{Gr01}.}  
 \label{f1}
\end{figure}
For the DWBA analysis of the considered \cx reactions, an accurate 
determination of the distorted waves in the entrance and exit channels 
by the appropriately chosen OP is very crucial. Although it is 
tempted to use consistently the total OP (\ref{e2}) given by the 
double-folding calculation using the same $t$-matrix interaction, such 
an attempt results on a poorer description of both the elastic scattering 
and \cx data. A plausible reason is that the higher-order medium corrections
to the microscopic OP are not negligible at the considered energy. Therefore,
we have used in the present DWBA analysis the phenomenological OP of the 
$^3$He+$^{90}$Zr system taken from Ref.~\cite{Kam03}. The complex OP of 
the $^3$He+$^{208}$Pb system has been obtained from a new OM fit \cite{Zer07} 
of the elastic $^3$He scattering data at 450 MeV \cite{Yam95}, with the 
relativistic kinematics. Following the earlier DWBA studies of the \he3t reactions 
\cite{Werf1,Werf2,Zer07,Per11}, the $^3$He OP rescaled by a factor $k=0.85$ has
been used fot the triton OP of the exit channel. All the DWBA calculations 
of the \cx scattering to the IAS were done with the relativistic kinematics,
using the code ECIS97 written by Raynal \cite{Ra97}.

\begin{figure}[bht] \vspace*{-1cm}\hspace{-1cm}
\includegraphics[width=1.0\textwidth]{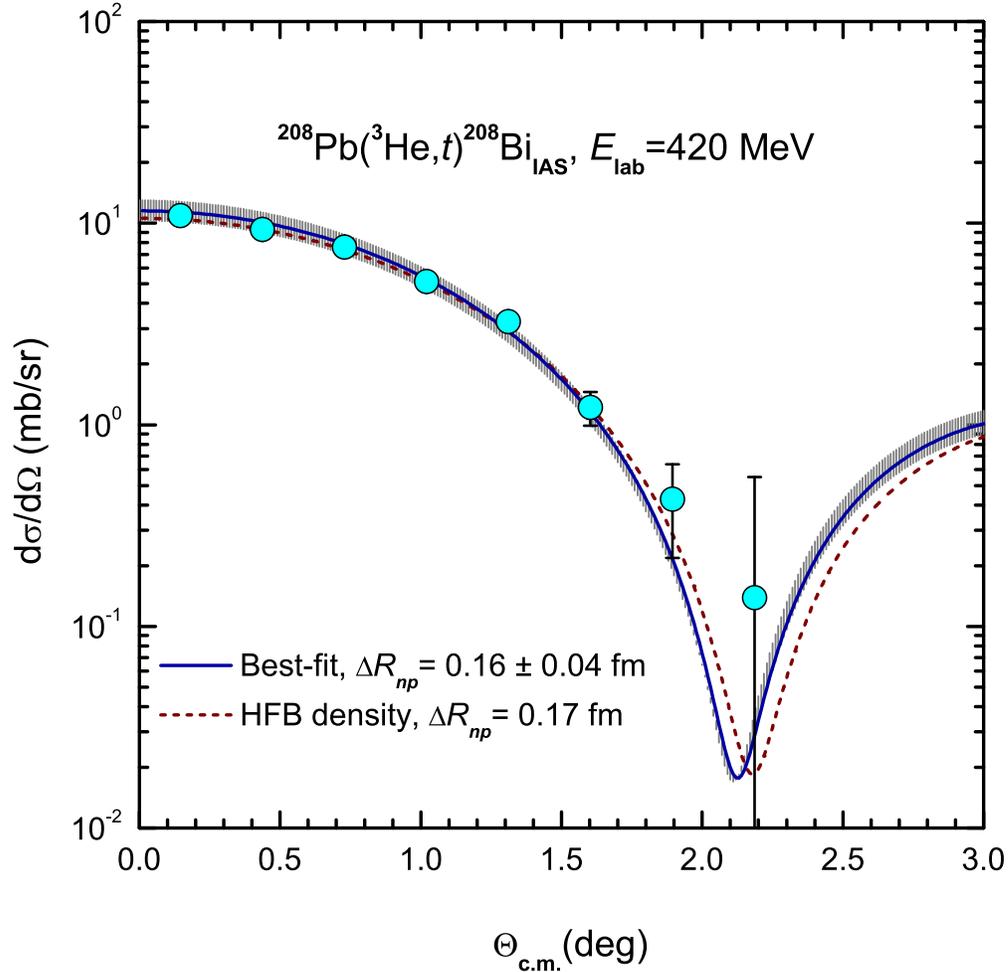}\vspace*{-1cm}
 \caption{The same as Fig.~\ref{f1} but for the  $^{208}$Pb target.}  
 \label{f2}
\end{figure}
The results of our DFM + DWBA analysis of the \he3t scattering to the IAS 
of the $^{90}$Zr target are shown in Fig.~\ref{f1}. Among the inputs of the
folding calculation of the \cx FF (\ref{e4}), only the radial parameter
of the empirical neutron density by Ray {\it et al.} \cite{Ray78} is slightly
adjusted to obtain the best DWBA fit to the \cx data. Such a simple linear 
fit resulted on a neutron density that gives the neutron skin 
$\Delta R_{np}\approx 0.09\pm 0.03$ fm. The uncertainty of the (best-fit) 
neutron skin is associated with the experimental uncertainty around 
10\% of the absolute differential cross section measured at the most 
forward angles. The obtained best-fit neutron skin is rather close to that
given by the microscopic HFB density ($\Delta R_{np}= 0.07$ fm) \cite{Gr01}. 
If the \cx FF (\ref{e4}) is calculated using the HFB density, then the DWBA 
results agree reasonably (within the error band) with those given by the 
(modified) empirical density. It is complimentary to note that the \pn data 
of the spin-dipole excitations of the $^{90}$Zr target have been analyzed 
\cite{Sag06} to deduce accurately the spin-dipole sum rule strength that gives 
the neutron skin $\Delta R_{np}\approx 0.07\pm 0.04$ fm for this nucleus, about
the same as that given by the HFB calculation. The best-fit neutron skin  
of $^{90}$Zr given by the present DFM + DWBA analysis is close to that 
($\Delta R_{np}\approx 0.085$ fm) given by the analysis of the elastic proton 
scattering data measured at $E_p=800$ MeV with the $^{90}$Zr target 
\cite{Ray78,Ray78c}.      

\begin{figure}[bht] \vspace*{-1cm}\hspace{-1cm}
\includegraphics[width=1.0\textwidth]{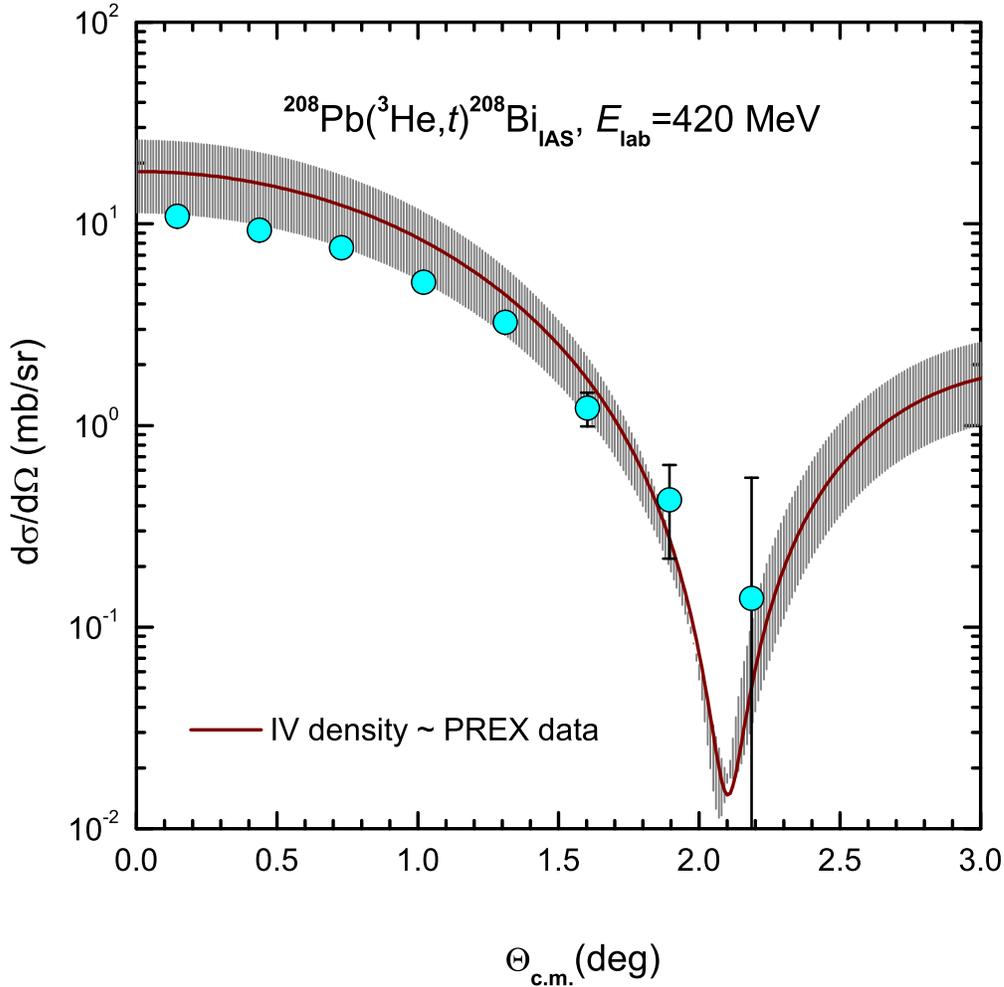}\vspace*{-1cm}
 \caption{DWBA description of the \he3t scattering to the IAS of the
$^{208}$Pb target given by the \cx FF (\ref{e4}) based on the empirical IV 
density by Ray {\it et al.} \cite{Ray78}, adjusted to reproduce the PREX data 
for the neutron skin, $\Delta R_{np}\approx 0.33^{+0.16}_{-0.18}$ fm \cite{prex}.}  
 \label{f3}
\end{figure}
Given the indirect relation of the neutron skin of $^{208}$Pb to the behavior 
of the nuclear symmetry energy, it has become a hot research topic recently   
\cite{Hor14,Tsang12,Roca13,prex,Fat13,Tar13}. Although the PREX data seem to  
provide for the first time an accurate, model-independent determination of  
the neutron skin of $^{208}$Pb \cite{prex}, the mean $\Delta R_{np}$ 
value deduced from the PREX data is significantly higher than that given 
by other studies \cite{Roca13,Fat13}, including the preliminary results 
of the $\gamma$-induced pion production \cite{Tar13}. The results of the 
DFM + DWBA analysis of the \he3t scattering to the IAS of $^{208}$Pb 
are shown in Fig.~\ref{f2}. After the radial parameter of the empirical
neutron density taken from Ref.~\cite{Ray78} was adjusted to the best 
DFM + DWBA fit to the \xpb data, a neutron skin 
$\Delta R_{np}\approx 0.16\pm 0.04$ fm has been obtained for $^{208}$Pb. 
As in the $^{90}$Zr case, the uncertainty of the best-fit $\Delta R_{np}$ 
value is resulted from the experimental uncertainty of about 10\% in the 
normalization of the absolute differential cross section measured 
at the forward angles. Although the error bars of the best-fit neutron
skin might be larger due to the uncertainty of the choice of the OP for the
entrance and exit channels, the $\Delta R_{np}$ value obtained in our
DFM + DWBA analysis is in a good agreement with that reported by other 
structure studies \cite{Tsang12,Roca13,Fat13,Tar13}. It is interesting
to note that the use of the microscopic HFB density \cite{Gr01} in the DFM 
calculation (with the associated $\Delta R_{np}=0.17$ fm) results on a very 
good overall agreement of the DWBA result with the \xpb data (see dash curve 
in Fig.~\ref{f2}).    

\begin{figure}[bht] \vspace*{-1cm}\hspace{-1cm}
\includegraphics[width=1.0\textwidth]{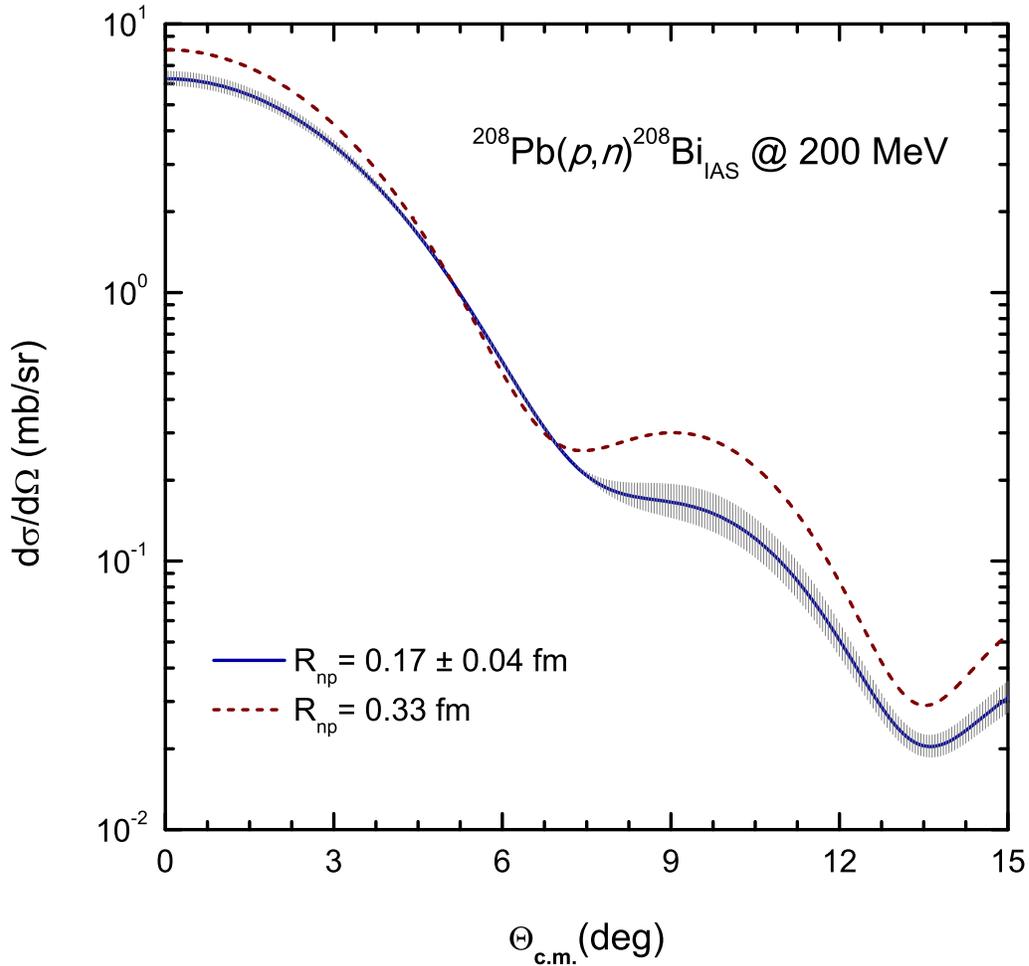}\vspace*{-1cm}
 \caption{DWBA prediction of the \pn scattering to the IAS of the
$^{208}$Pb target at $E_p=200$ MeV given by the folded FF based on the 
empirical IV density by Ray {\it et al.} \cite{Ray78}, adjusted to reproduce 
the best-fit neutron skin of $^{208}$Pb given by the present DFM + DWBA analysis 
of the \he3t data and the mean $\Delta R_{np}$ value given by the PREX data 
\cite{prex}.}  \label{f4}
\end{figure}
For a comparison, we have made further a DFM + DWBA calculation using the 
IV nuclear density of $^{208}$Pb constructed to give the same neutron 
skin as that given by the PREX data. From these results (see Fig.~\ref{f3}) 
one can see that the lower edge of the PREX data agrees nicely with 
the measured \he3t data. Consequently, we still cannot rule out the large 
neutron skin of $^{208}$Pb given by the PREX measurement, as was also 
concluded recently by Fattoyev and Piekarewicz \cite{Fat13}. The new PREX
experiment planned to pin down the uncertainty of the $\Delta R_{np}$
value to about 0.06 fm \cite{Hor14} would surely resolve the uncertainty 
of the DFM + DWBA results shown in Fig.~\ref{f3}.

Despite numerous elastic proton scattering data taken at medium energies, 
the high-precision data of the \pn scattering to the IAS at medium energies are 
still not available. We found it of interest to make a folding + DWBA 
prediction of the \pn scattering to the IAS of $^{208}$Pb at the proton energy 
of 200 MeV. From the results shown in Fig.~\ref{f4} one can see that the effect
caused by different neutron skin values is significantly not only at the zero
scattering angle but also at the first diffraction maximum. With the modern
neutron detection technique, it should be feasible to cover the whole 
oscillation pattern of the \pn scattering cross section at the forward
region and, eventually, allow one to fine tune the neutron skin value in
a similar folding + DWBA analysis.  
      
In conclusion, the existing data of the \he3t scattering to 
the IAS of $^{90}$Zr and $^{208}$Pb at $E_{\rm lab}=420$ MeV have been studied 
in a detailed DFM + DWBA analysis to deduce the neutron skin values for these 
two nuclei. The best-fit neutron skin values given by our analysis are in a 
good agreement with those given by the recent nuclear structure studies. 

The present research has been supported, in part, by the National Foundation 
for Science and Technology Development (NAFOSTED project No. 103.04-2011.21), 
by the LIA program of the Ministry of Science and Technology of Vietnam, 
and by the US NSF (PHY-1102511). We also thank Eduardo Garrido 
and Marcella Grasso for providing the microscopic nuclear densities 
for the DFM calculation.

\end{document}